\documentstyle[aps,epsfig]{revtex}

\def  \uG     {\underline{G}}

\def  \tsig   {\tilde{\sigma}}
\def  \bnab   {{\boldsymbol \nabla}}
\def  \bsig    {\mbox{\boldmath$\sigma$}}
\def  \balph   {\mbox{\boldmath$\alpha$}}
\def  \bnab    {\mbox{\boldmath$\nabla$}}

\def  \epsi     {\varepsilon}

\def  \eff    {\mathit{{\mathrm eff}}}
\def  \op     {{\bf p}}

\begin{document}

\title{Anomalous Hall effect and weak localization
corrections in a ferromagnet}
\author{A.~Cr\'epieux$^a$, J.~Wunderlich$^a$, V.K.~Dugaev$^{a,b}$ and P.~Bruno$^a$}
\address{$^a$Max-Planck-Institut f\"{u}r Mikrostrukturphysik, Weinberg
2, 06120 Halle, Germany}
\address{$^b$Institute of Materials Science Problems, Vilde~5, 58001 Chernovtsy,
Ukraine}
\date{\today}
\maketitle

\begin{abstract}
In this paper, we report results on the anomalous Hall effect.
First, we summarize analytical calculations based on the Kubo
formalism : explicit expressions for both skew-scattering and
side-jump are derived and weak-localization corrections are
discussed. Next, we present numerical calculations of the
anomalous Hall resistivity based on the Dirac equation.
Qualitative agreement with experiments is obtained.\\
\end{abstract}

The anomalous Hall resistivity corresponds to the spontaneous
value that takes in magnetic materials the Hall resistivity in the
absence of applied magnetic field and results from a combination
of spin-orbit coupling and spin polarization. Two different
mechanisms responsible for this effect are distinguished: the
skew-scattering \cite{Smit} and the side-jump \cite{Berger}, both
corresponding to an asymmetric spin-dependent scattering by a
potential in the presence of the spin-orbit coupling. The
simplest way to calculate such an effect is to start from the
Pauli equation including the spin-orbit coupling
\begin{eqnarray}\label{Pauli}
  H=\frac{p^2}{2m}+V-\mu_B(\bsig\cdot{\bf B}_{\eff})+\frac{\hbar}{4m^2c^2}(\bsig\times\bnab V)\cdot{\bf p},
\end{eqnarray}
and to use the Kubo formula for the conductivity
\begin{eqnarray}\label{Kubo}
  \tsig_{ij}=\frac{e^2\hbar}{2\pi\Omega}
  {\mathrm Tr}\left\langle v_iG^+(\epsi_F)v_jG^-(\epsi_F)\right\rangle_c,
\end{eqnarray}
where $\Omega$ is the volume of the sample, $\langle...\rangle_c$
denotes the configurational average, $G^+$ and $G^-$ are the
retarded and advanced Green's functions
$G^{\pm}(\epsi)=\left(\epsi\pm i0-H\right)^{-1}$, $\epsi_F$ is
the Fermi level and $v_i$ is the i-component of the velocity
which contains an additional part due to spin-orbit coupling (the
so-called anomalous velocity ${\bf v}_{SO}$)
\begin{eqnarray}\label{Velocity}
  {\bf v}=\frac{\op}{m}+{\bf v}_{SO}=\frac{\op}{m}+\frac{\hbar}{4m^2c^2}(\bsig\times\bnab V).
\end{eqnarray}
The anomalous velocity inserted in (\ref{Kubo}) gives the
side-jump term whereas the spin-orbit coupling contribution to
the Green's function (i.e., $G_0H_{SO}G_0$ where $H_{SO}$ is the
spin-orbit coupling and $G_0$ the Green's function associated to
the Hamiltonian in the absence of the spin-orbit coupling)
inserted in (\ref{Kubo}) gives the skew-scattering term provided
one goes beyond the Born approximation \cite{Smit}.

We present a simple application of such a method of calculation
in the case of a system with a cubic symmetry and an effective
magnetic field along the z-axis: thus the anomalous Hall
conductivity is equal to the off-diagonal element $\tsig_{xy}$.
We model the compound in the following way: the total volume of
the sample $\Omega$ is divided into $N$ cells of volume
$\Omega_0$. In each cell, the potential takes a constant value
$V$ with a probability distribution $P(V)$ which is characterized
by its moments $\langle V^n\rangle_c = \int P(V)V^ndV$. A suitable
choice of the energy origin yields $\langle V\rangle_c=0$. We
assume that there are no correlations in the value of the
potential in different cells. In order to achieve analytical
calculations, we restrict the study to the lowest order with the
scattering potential and express the Green's function $G$ in term
of the average Green's function $\uG$ in the relaxation time
approximation. Neglecting weak-localization corrections, the
diagonal conductivity is then given by the Einstein relation and
for the off-diagonal conductivity, we get
\begin{eqnarray}\label{Skew-scattering}
  \tsig_{xy}^{SS}
  =-\frac{\pi m^2\lambda^2}{6\hbar^2}\frac{\langle
  V^3\rangle_c}{\langle V^2\rangle_c}
  \left({\cal N}_{\uparrow}\Omega_0\tsig_{xx}^{\uparrow}(v_F^{\uparrow})^2
  -{\cal N}_{\downarrow}\Omega_0\tsig_{xx}^{\downarrow}(v_F^{\downarrow})^2\right)
\end{eqnarray}
for the skew-scattering, and
\begin{eqnarray}\label{Side-jump}
  \tsig_{xy}^{SJ}=-e^2{\cal N}_{\uparrow}\frac{2\delta^{\uparrow}v_F^{\uparrow}}{3}+e^2{\cal
  N}_{\downarrow}\frac{2\delta^{\downarrow}v_F^{\downarrow}}{3}
\end{eqnarray}
for the side-jump, respectively (detailed calculations are
presented in Ref.~\cite{Crepieux}). We have introduced the
quantities $\lambda=\hbar/mc$,
$\delta^{\uparrow(\downarrow)}=\hbar
v_F^{\uparrow(\downarrow)}/4mc^2$. ${\cal
N}_{\uparrow(\downarrow)}$, $v_F^{\uparrow(\downarrow)}$ and
$\tsig_{xx}^{\uparrow(\downarrow)}$, are respectively, the density
of states per unit volume $\Omega_0$, the velocity at Fermi energy
and the diagonal conductivity, each for up and down spins. In
contrast to $\tsig_{xx}$ and $\tsig_{xy}^{SS}$, $\tsig_{xy}^{SJ}$
does not depend on disorder. The Feynmam diagrams associated with
these mechanisms are depicted in Fig.~(1). A simple illustration
of these results can be given in the case of a binary alloy
${\mathrm A_xB_{1-x}}$ for which we have $\langle
V^2\rangle_c=x(1-x)(\epsi_A-\epsi_B)^2$ and $\langle
V^3\rangle_c=x(1-x)(1-2x)(\epsi_A-\epsi_B)^3$ where $\epsi_{A(B)}$
is the value that takes the potential on side A(B). As a
consequence, the anomalous Hall resistivity for skew-scattering
is equal to
$\tilde{\rho}_{H}^{SS}\simeq\tsig_{xy}^{SS}/\tsig_{xx}^2\propto
(x-3x^2)$ and for side-jump to
$\tilde{\rho}_{H}^{SJ}\simeq\tsig_{xy}^{SJ}/\tsig_{xx}^2\propto
x^2$ which is in agreement with the empirical relation
$\tilde{\rho}_{H}=a\tilde{\rho}_{xx}+b\tilde{\rho}^2_{xx}$ but in
disagreement with the common belief that the quadratic term arises
only from the side-jump.

Within this approach, the weak-localization corrections to the
anomalous Hall effect can also be calculated. We have considered
both Cooperons and Diffusons (see Fig.~(1)). The results are the
following \cite{Dugaev}: (i) the Cooperons diagrams for the
side-jump cancel exactly each other whereas the Diffusons
diagrams give a negligible contribution (of order
$\left(\hbar/\epsi_F\tau\right)^4$ where $\tau$ is the relaxation
time), (ii) the Cooperons diagrams for the skew-scattering give a
non-zero contribution which includes both spin up and spin down
channels. As a consequence, the weak-localization corrections to
the anomalous Hall resistivity
$\Delta\tilde{\rho}_H/\tilde{\rho}_H^0\simeq
\Delta\tsig_{xy}/\tsig_{xy}^0-2\Delta\tsig_{xx}/\tsig_{xx}^0$
exhibit a strikingly different behavior as compared to the normal
Hall resistivity for which weak-localization corrections vanish by
an exact cancellation of the diagonal and off-diagonal parts
\cite{Fukuyama}. Due to the presence of two spin channels, such a
cancellation can never take place in the case of the anomalous
Hall resistivity.

The analytical expressions (\ref{Skew-scattering}) and
(\ref{Side-jump}) have been obtained in the weak-scattering limit
and using the free electron approximation. In order to get a more
realistic description of the anomalous Hall effect, we have
performed numerical calculations starting from a tight-binding
description and using the coherent potential approximation in
order to treat disorder. The Green's function is then expressed
in term of the t-matrix and the average Green's function: $G=\uG
+\uG T\uG $. As a consequence, the conductivity (\ref{Kubo}) can
be split, in the case of a configuration-independent velocity,
into two different parts
\begin{eqnarray}\label{Vertex}
  \tsig_{ij}=\frac{e^2\hbar}{2\pi\Omega}
  {\mathrm Tr}\left[v_i\uG^+(\epsi_F)v_j\uG^-(\epsi_F)\right]
  +\frac{e^2\hbar}{2\pi\Omega}{\mathrm Tr}\left[v_i\uG^+(\epsi_F)\Gamma_j(\epsi_F)\uG^-(\epsi_F)\right],
\end{eqnarray}
where $\Gamma_j$ is the vertex function equal to $\langle
T^+\uG^+v_j\uG^-T^-\rangle_c$. This formulation is very useful
since it allows the exact determination of the vertex corrections
in the ladder approximation. However, because of the presence of
the spin-orbit coupling, the velocity is no longer
configuration-independent (see Eq.~(\ref{Velocity})) and it is
then not possible to take it out of the configuration average
$\langle...\rangle_c$ like it is done in Eq.~(\ref{Vertex}). A
means to avoid this problem is to start, rather than from the
Pauli equation, from the Dirac equation
\begin{eqnarray}\label{Dirac}
  H=c\left(\balph\cdot\op\right)+\beta
  mc^2+V-\mu_B\beta\left(\bsig\cdot{\bf B}_{\eff}\right),
\end{eqnarray}
where $\balph$ and $\beta$ correspond to the standard Dirac
matrices. Indeed, in this description, the velocity operator is
simply equal to $c\balph$ and then is configuration-independent.

The system we consider is a ferromagnetic binary alloy ${\mathrm
A_xB_{1-x}}$ with an effective magnetic field along the z-axis.
First, we have expressed Eq.~(\ref{Dirac}) in the tight-binding
approximation for a cubic symmetry, next the self-energy has been
numerical calculated by the means of an iteration procedure and
used to calculate the average Green's function and the vertex
corrections that we insert in Eq.~(\ref{Vertex}) in order to get
the conductivity tensor. Variations of the anomalous Hall
resistivity $\tilde{\rho}_{H}\simeq\tsig_{xy}/\tsig_{xx}^2$ as a
function of concentration of disorder and spin-orbit coupling are
depicted in Fig.~(2). We obtain a change of sign of the anomalous
Hall resisitivity for a particular value of the concentration
which corresponds to an exact cancellation of the skew-scattering
and the side-jump contributions. Such a change of sign has been
observed experimentally in PdCo and PdNi alloys \cite{Hamzic}.
The variation of the anomalous Hall resistivity with spin-orbit
coupling is consistent with the Onsager relation since the
numerical curve can be very well fitted when one consider only
odd powers with respect to the spin-orbit coupling. A very good
agreement with analytical expressions which apply in the
weak-scattering limit (close to $x=0$ and $x=1$) and in the
weak-relativistic limit ($\lambda_{SO}\simeq0$) is obtained.

To summarize, we have performed both analytical and numerical
calculations concerning the anomalous Hall effect. Explicit
expressions of the skew-scattering and side-jump conductivities
have been derived which allows the clarification of the influence
of the potential on each mechanisms: whereas the skew-scattering
conductivity varies on the third order of the potential, the
side-jump conductivity does not depend on it. Contrary to what
happens for the normal Hall resistivity, the weak-localization
corrections to the anomalous Hall resistivity do not vanish
because of the presence of two different spin channels. Numerical
results obtained in the case of a ferromagnetic binary alloy are
in qualitative agreement with measurements in the sense that they
show a change of sign of the anomalous Hall resistivity.

V.D. is thankful to J. Barna\'s for numerous discussions. Part of
this work is supported by KBN Grant No.~5~P03B~091~20 and NATO
Linkage Grant  No.~977615.

\begin{figure}
\epsfig{file=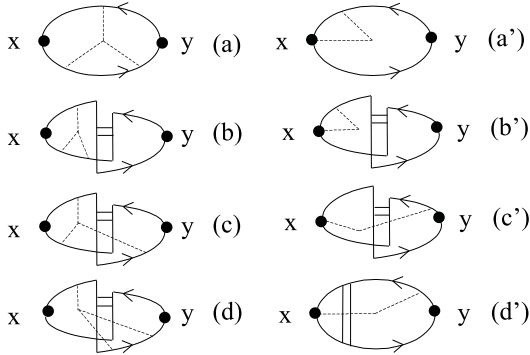} \caption{Feynman diagrams for the
anomalous Hall effect. (a) corresponds to the skew-scattering,
(a') to the side-jump. (b), (c) and (d) are the Cooperons
corrections to the skew-scattering, (b') and (c') to the
side-jump. (d') is the Diffusons correction to the side-jump. The
curve lines represent the average Green's functions, the dashed
lines correspond to the potential (including the spin-orbit
coupling) and the double straight lines correspond to the ladder
part. Symmetrical diagrams have also to be considered.}
\end{figure}

\begin{figure}
\epsfig{file=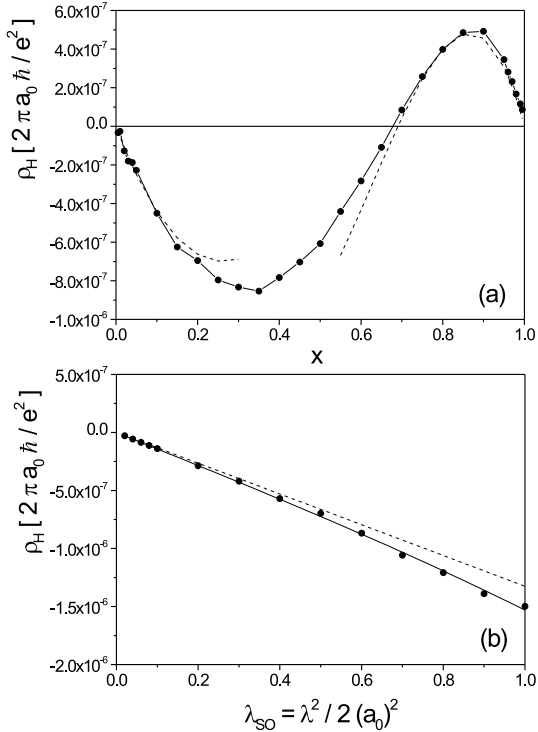} \caption{Variation of the anomalous
Hall resistivity as a function (a) of the concentration for a
spin-orbit coupling $\lambda_{SO}=0.5$ and as a function (b) of
the spin-orbit coupling for a concentration $x=0.2$. The others
parameters are $\epsi_F/t=0.2$, $\epsi_A/t=0.1$ and $\epsi_B/t=0$
where $t=\hbar/2ma_0^2$ ($a_0$ is the unit cell parameter) and
$\Delta\epsi_A/t=\Delta\epsi_B/t=0.05$ where
$\pm\Delta\epsi_{A(B)}/2$ is the value that takes the exchange
coupling $-\mu_B\sigma_zB_{\eff}$ on site A(B). The full lines
with full circle symbols correspond to the numerical calculations
and the dashed lines correspond to the analytical results given by
Eqs.~(\ref{Skew-scattering}) and (\ref{Side-jump}).}
\end{figure}

\end{document}